\documentclass[superscriptaddress,aps,prl,twocolumn,preprintnumbers,amsmath,amssymb,floatfix,groupeda
ddress]{revtex4}
\usepackage{graphicx}
\begin{document}
\title{Highly Efficient Coherent Optical Memory Based on Electromagnetically Induced Transparency}
\author{Ya-Fen Hsiao$^{1,2}$, Pin-Ju Tsai$^{1,3}$, Hung-Shiue Chen$^1$, Sheng-Xiang Lin$^{1,3}$, Chih-Chiao Hung$^1$, Chih-Hsi Lee$^1$, Yi-Hsin Chen$^4$, Yong-Fan Chen$^5$, Ite A. Yu$^{4,*}$, and Ying-Cheng Chen}
\email{yu@phys.nthu.edu.tw; chenyc@pub.iams.sinica.edu.tw}
\affiliation{Institute of Atomic and Molecular Sciences, Academia Sinica, Taipei 10617, Taiwan
\\
$^{2}$Molecular Science and Technology Program, Taiwan International Graduate Program, Academia Sinica and National Central University, Taiwan 
\\
$^{3}$Department of Physics, National Taiwan University, Taipei 10617, Taiwan
\\
$^{4}$Department of Physics, National Tsing Hua University, Hsinchu 30043, Taiwan
\\
$^{5}$Department of Physics, National Cheng Kung University, Tainan 70101, Taiwan}
\date{\today}
\begin{abstract}
Quantum memory is an important component in the long-distance quantum communication system based on the quantum repeater protocol. To outperform the direct transmission of photons with quantum repeaters, it is crucial to develop quantum memories with high fidelity, high efficiency and a long storage time.  Here, we achieve a storage efficiency of 92.0(1.5)\% for a coherent optical memory based on the electromagnetically induced transparency (EIT) scheme in optically dense cold atomic media. We also obtain a useful time-bandwidth product of 1200, considering only storage where the retrieval efficiency remains above 50\%. Both are the best record to date in all kinds of the schemes for the realization of optical memory. Our work significantly advances the pursuit of a high-performance optical memory and should have important applications in quantum information science.
\end{abstract}
\maketitle
Quantum memory is a device that can store and then retrieve a quantum state on demand\cite{Tittel13}. It is an important building block in quantum communication\cite{DLCZ,Gisin11,Tittel13} and quantum computation\cite{Kok07}. Parameters for evaluating the performance of a quantum memory include the fidelity, efficiency, storage time, capacity and bandwidth\cite{Tittel13}. In the past few years, significant progress has been made in improving these performance parameters\cite{Guo12,Buchler11,Buchler16,Kuzmich13,Halfmann13,Guo13,Walmsley11}.
In this work, we focus our discussion on the efficiency. Quantum memory with a high efficiency is crucial in many applications, such as quantum network\cite{Kimble08}, many-photon synchronization\cite{Nunn13}, and the quantum repeater for long-distance cryptography\cite{DLCZ,Gisin11}. Taking quantum repeater as an example, an increase of 1\% in efficiency may decrease the entanglement distribution time by 7-18\%, depending on the protocol\cite{Gisin11,Tittel13}. This highlights the importance to achieve a high efficiency.

The memory efficiency is determined both by technical losses and the storage efficiency (SE), defined by the ratio of the energy of the stored-and-retrieval pulse to that of input pulse in the absence of gain or other added noise. SE evaluates the intrinsic efficiency of a memory, which depends on the storage mechanism and the media property. Many memory protocols have been proposed and implemented, including off-resonant Raman interactions\cite{Walmsley11}, atomic frequency combs (AFC)\cite{Gisin10, Gisin13}, gradient echo\cite{Buchler11,Buchler16}, and electromagnetically induced transparency (EIT)\cite{Lukin00,Novikova07,Novikova08,Lvovsky08,Lvovsky09,Du11,Du12,Yu13,Schraft16}.   
Under the demonstration of quantum storage, some impressive storage efficiencies attained are 87\% using the gradient echo in warm atomic vapors\cite{Buchler16} and 69\% in solid-state medium\cite{Hedges10}, and 68\% using EIT with cold atoms\cite{Laurat18}. In the regime of classical storage, a SE of 78\%\cite{Yu13} and 76\%\cite{Schraft16} have been achieved using EIT in cold atoms and solid-state medium, respectively. In this work, we report a SE of 92.0(1.5)\% using EIT with cold atoms in the classical storage.   

Since a high SE is crucial in memory applications, it is essential to explore what mechanisms limit one from getting a high SE. A high optical depth (OD, denoted as $D$) and a low ground-state decoherence rate ($\gamma_{21}$) for the media are the basic requirements to obtain a high SE for EIT-based memory\cite{Lukin07,Lukin07b}. However, at high ODs, some nonlinear optical effects may become significant to induce complications. For example, the off-resonant excitation of the control field on a nearby excited state causes a multiphoton decay channel\cite{Schmidt96,Harris98}, and induces a control-intensity-dependent $\gamma_{21}$ which leads to a degraded SE. By choosing $D_1$-line to implement the EIT, we resolve this problem and reach a high SE which is impossible to obtain with the $D_2$-line system\cite{Yu13}. At high ODs, the off-resonant excitation of the control field on the probe transition induces the four-wave mixing (FWM) \cite{Lukin88,Lett07,Novikova09,Fleischhauer13,Boyer13} which leads to a probe gain and a degradation in memory fidelity\cite{Fleischhauer13}. Among the stable alkalis, cesium has the largest ratio of the ground-state hyperfine splitting to the excited-state spontaneous decay rate which favors a minimal FWM effect\cite{Fleischhauer13}. In addition, one can reduce the FWM by slightly misaligning the beams to break the phase matching condition. 

Based on these, we have achieved a SE of 92.0(1.5)\% with an experimental check on FWM gain of $<$3\%. This SE is the best record to date in all kinds of memory scheme. Although our experiment were done with coherent probe pulses containing $\sim$ 30000 photons, the result could be extended to the quantum regime\cite{Lvovsky08,Lvovsky09,Pan11,Du11,Du12}. In a memory application, the time-bandwidth product (TBP), defined as the ratio of the storage time at 50\% SE to the FWHM input pulse duration ($T_p$), is another crucial figure of merit which evaluates how many operations a memory can provide. We have achieved a record-high TBP of 1200. Compared to\cite{Yu13}, we use a shorter $T_p$ of 200 ns that favor this product. Another advantage of using a shorter $T_p$ is that the SE is less susceptible to the reduction by a finite $\gamma_{21}$. The simultaneous high SE and TBP is important in memory-based applications such as many-photon synchronization\cite{Nunn13}.  

EIT-based memory relies on the slow light effect or the dark-state polariton which is a coherent superposition of the optical field and the collective atomic excitation. By adiabatically ramping the control field off and on, the optical component can be coherently converted into atomic component and stored inside the media until being retrieved as an optical field\cite{Lukin00}. The overall SE $\eta$ depends on three factors, denoted as $\eta_{tran}$, $\eta_{comp}$ and $\eta_{stored}$, such that $\eta=\eta_{tran}\eta_{comp}\eta_{stored}$\cite{supp}. $\eta_{tran}$ is the transmission of a slow light pulse. $\eta_{comp}$ indicates the fraction of a probe pulse that is compressed into the medium during storing. $\eta_{stored}$ indicates the remaining efficiency after a given storage time. For a large enough OD (see\cite{supp}) and a short storage time, both $\eta_{comp}$ and $\eta_{stored}$ can be close to one and
$\eta_{tran}$ dominates the SE.

Based on Maxwell-Bloch equations, one can show that the transmission of a Gaussian probe pulse with an intensity FWHM width of $T_p$ is \cite{Chen11,supp},
\begin{equation}\label{eta} 
\eta_{tran}=\frac{e^{-2\gamma_{21} T_d}}{\sqrt{1+32ln2\frac{\gamma_{31}}{\Gamma}\frac{\zeta^2}{D}}}=\frac{e^{-2\gamma_{21} T_d}}{\sqrt{1+(\frac{4ln2}{T_p\Delta\omega_{EIT}})^2}},
\end{equation} 
where $\Delta\omega_{EIT}\cong\sqrt{\frac{ln2}{2}}\frac{\Omega_c^2}{\sqrt{D\gamma_{31}\Gamma}}$ is the FWHM EIT bandwidth, $\Omega_c$ is the Rabi frequency of the control field, $T_d\cong\frac{D\Gamma}{\Omega_{c}^{2}}$ is the group delay, $\zeta\equiv\frac{T_d}{T_p}$, $\gamma_{31}$ is the decay rate of the optical coherence $\rho_{31}$, and $\Gamma$ is the spontaneous decay rate, which is $2\pi\times$4.575(5.234) MHz for cesium 6$P_{1/2}$ (6$P_{3/2}$) state\cite{Gould03}. Eq.(1) indicates that $\eta_{tran}$ is limited by the finite $\gamma_{21}$ and the finite EIT bandwidth. A small $\gamma_{21}$ and a large OD are two key parameters to obtain a high SE at a fixed $\zeta$. To keep $\zeta$ fixed at a given $T_p$ for a larger OD, one has to increase the $\Omega_c$. In the ideal case with $\gamma_{21}=0$ and $\gamma_{31}=\Gamma/2$,  $\eta_{tran}$ approaches unity with the scaling law $\eta_{tran}=1-\frac{40}{D}$. Here, we assume $\zeta=2.7$ such that $\eta_{comp}>$0.99\cite{supp}. In Refs.\cite{Lukin07,Lukin07b}, the authors describe an optimal method in which the probe waveform is optimized to maximize the SE. Compared to the case of a Gaussian waveform, this method gains a significant improvement in SE at moderate ODs but gains a little at high ODs. A similar scaling law was derived\cite{Lukin07b}. 

\begin{figure}
\includegraphics[width=8 cm,viewport=160 60 690 550,clip]{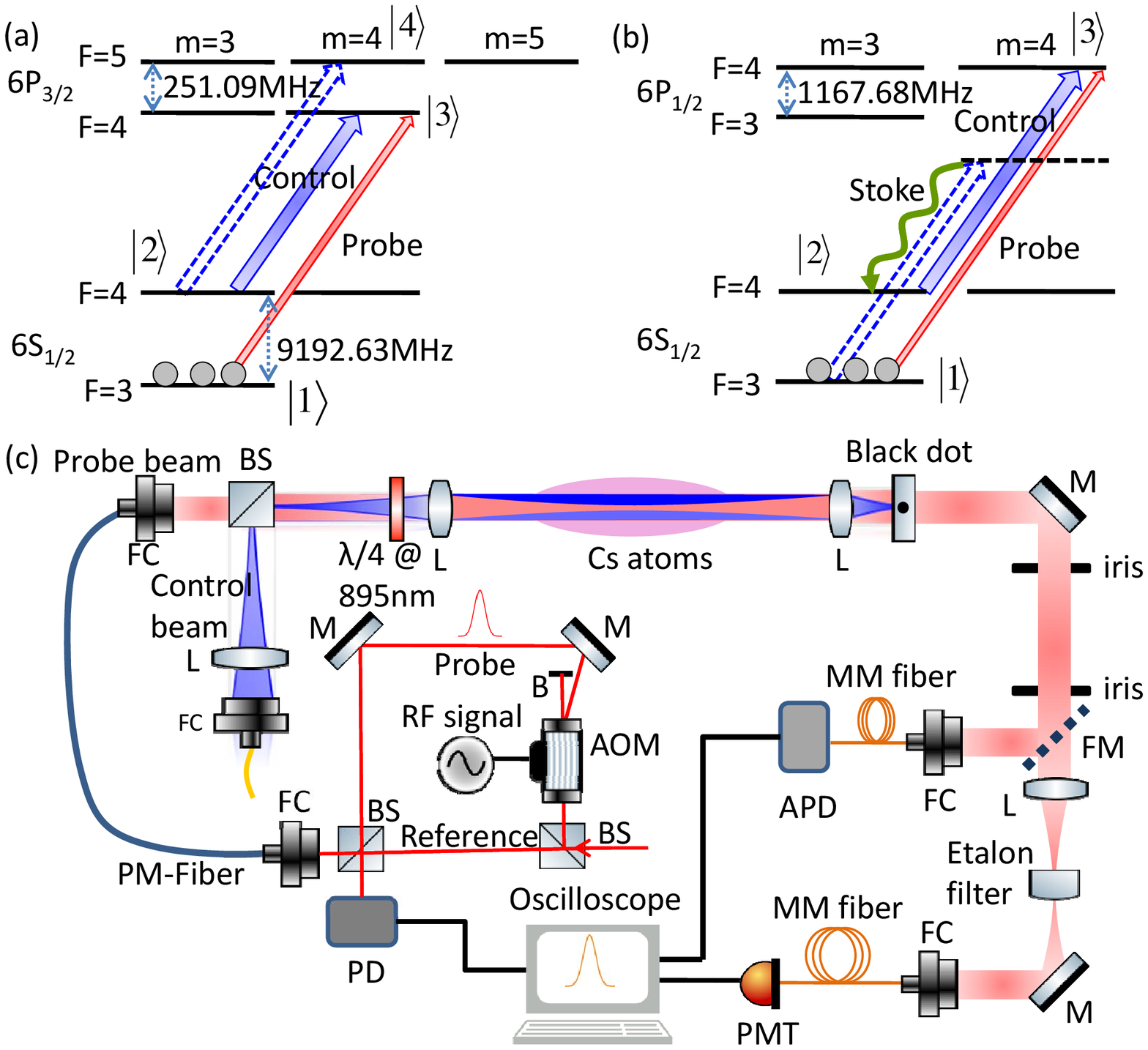}
\caption{(a) and (b): Relevant energy levels and laser excitations for the $D_2$ and $D_1$ EIT scheme, respectively. (c) Schematic experimental setup. AOM: acousto-optic modulator; APD: avalanche photodetector; B: block; BS: beam splitter; FC: fiber coupler; FM: flip mirror; L: lens; M: mirror; MM fiber: multimode fiber; PD: photodetector; PM fiber: polarization-maintaining fiber; PMT: photomultiplier tube.} 
\end{figure}  

We utilize a magneto-optical trap of cesium in addition to many techniques to obtain cold atomic medium with ODs of up to $\sim$1000\cite{Chen14b}. 
Many efforts are made to reduce the decoherence rate ($\gamma_{21}$) to $10^{-4} \Gamma$ level. The linewidth of the beatnote of the control and probe lasers is kept to less than 10 Hz by injection locking both lasers to a master laser. The linewidth of the master laser is measured to be less than 100 kHz by the delayed self-heterodyne method\cite{Richter86}. The overall dc and ac stray magnetic fields are reduced to 5 mG level. The control and probe beams are copropagating with an angle $\theta$ of $<1^0$ to reduce the decoherence due to atomic motions\cite{Pan08,Yu11}. Under such a condition, leakage of the control light into the detector for probe detection becomes an issue. A series of arrangements including a window with a black dot, some irises, an etalon filter (Quantaser FPE001) and a multimode fiber are used to obtain a 73 dB isolation of the control power (Fig. 1(c)). We also use the beatnote interferometer to check the coherence property of the slow and stored pulses\cite{Yu05}. More details are  shown in\cite{supp}.  

We implemented the EIT-based memories with both cesium $D_2$ and $D_1$ lines with the level schemes shown in Fig.1 (a) and (b). The population is mainly prepared in the $|F=3,m=3\rangle$ Zeeman state by the optical pumping\cite{Chen14b}. With $\sigma^+$-polarized control and probe fields, the EIT system mainly involves three Zeeman sublevels. However, the control field can off-resonantly couple to the nearby $|2\rangle\rightarrow|4\rangle$ transition in the $D_2$ scheme. This coupling induces an additional channel for multiphoton loss\cite{Yu13}, which is called the N-type photon switching effect\cite{Schmidt96,Harris98}. It introduces a $\Omega_c$-dependent $\gamma_{21}$ and limits the SE at high ODs\cite{Yu13}. Instead, the $D_1$ scheme is free from such a complication because there is no any nearby excited state. To illustrate this effect, we take EIT spectra for various control intensities at a given OD for both schemes. A significant difference in the degree of transparency at EIT resonances is observed between the two schemes\cite{supp}. The transparency is only up to 70\% for the $D_2$ scheme but is almost 100\% for the $D_1$ scheme. By fitting the spectra to the corresponding lineshape, the decoherence rate $\gamma_{21}$ can be obtained\cite{supp}. Fig. 2(a) depicts $\gamma_{21}$ versus the control power for both schemes. As expected, $\gamma_{21}$ for the $D_2$ scheme are much larger than those of $D_1$ scheme and scales linearly with the control power. Therefore, $D_1$ scheme is a better choice and we focus our study on this scheme. The off-resonant coupling of the control field also causes an ac Stark shift, as shown in Fig. 2(b). More details are shown in\cite{supp}.
\begin{figure}
\includegraphics[width=8.5 cm,viewport=225 180 780 365,clip]{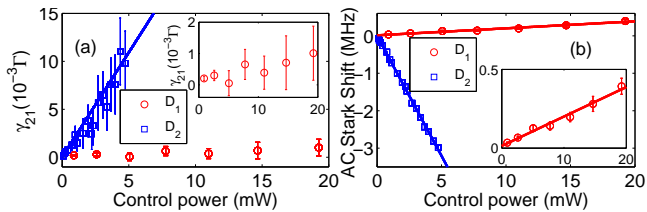}
\caption{(a) $\gamma_{21}$ and (b)AC Stark shift versus the control power for both $D_1$ and $D_2$ EIT schemes. The insets are the y-scale zoom of the $D_1$ data. In (a), the blue line is a calculation with $\gamma_{21}=\gamma_0+\frac{(\sqrt{48/7}\Omega_c)^2}{4\delta_s^2}\gamma_{41}$, where $\gamma_0=0.0001\Gamma$, $\gamma_{41}=0.8\Gamma$. In (b), the red (blue) line is a linear fit to the $D_1$ ($D_2$) data. The slope is positive (negative) for the $D_1$ ($D_2$) system because the control field causes an AC Stark shift on state $|1\rangle$ ($|2\rangle$) due to the red-detuned, off-resonant excitation of the probe ($|2\rangle$ $\rightarrow$ $|4\rangle$) transition.}
\end{figure}   

Fig. 3(a) depicts a representative EIT spectrum of the $D_1$ scheme. In this case, the fitting parameters \{$D$, $\Omega_c$, $\gamma_{21}$, $\delta_c, \gamma_{31}$\} are \{822(53), 7.41(4)$\Gamma$, 0.0004(4)$\Gamma$, -0.012(6)$\Gamma, 1.07(12)\Gamma$\}, respectively. Quantities in the brackets are 2$\sigma$ standard deviations. These parameters are determined by the joint fitting of both EIT spectrum and the slow light trace\cite{supp}. We found that the obtained $\gamma_{31}\simeq 0.7\Gamma$ at low OD and increases as OD increases\cite{supp}. The deviation from the ideal case of $0.5\Gamma$ for $\gamma_{31}$ may due to the finite laser linewidth and laser frequency fluctuations. The spectral broadening as OD increases may due to the cooperative effect by resonant dipole-dipole interactions\cite{Jennewein16}, which certainly deserves a further study but is not the focus of this work. Fig. 3(b) depicts one representative dataset of the input, slow and stored-and-retrieved probe pulses taken under the same condition as that of Fig. 3(a). The efficiency of the slowed and stored pulse is 92.0(1.4)\% 91.2(1.1)\%, respectively. The purple line in Fig. 3(b) is a calculated slow light trace with the experimentally determined parameters. Fig. 3(c) depicts the representative beatnote data for the input, slow, and stored-and-retrieved pulses. Fig. 3(d)-(e) show parts (50 ns duration) of the data around the peaks of the three traces. Quantitatively, the classical fidelities which characterize the resemblance of the electric field of the slow or retrieved pulse to the input one\cite{Yu13,Yu13b} are 98.1\% and 93.1\%, respectively. This reflects that phase coherences are well preserved for both the slow and retrieved pulses.

\begin{figure}
\includegraphics[width=8 cm,viewport=235 5 795 450,clip]{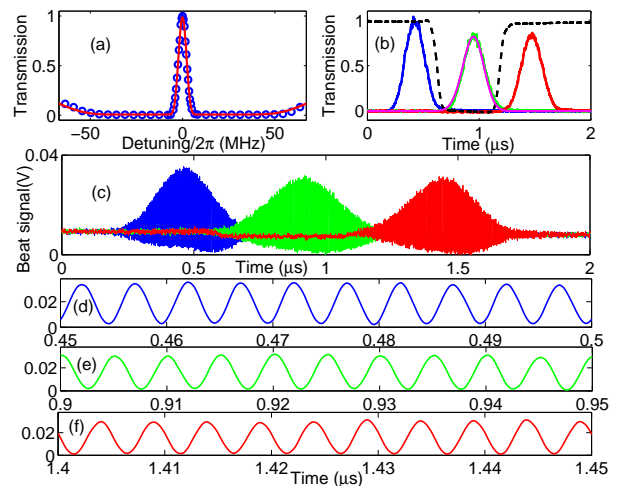}
\caption{(a) A representative EIT spectrum. The red line in a fitting curve. (b) Blue, green and red traces are the input, slow, and stored-and-retrieval pulse, respectively. The dashed black trace is the control intensity. The purple line is a calculated slow light curve. (c) The representative beatnote data.(d),(e),(f) Parts of the beatnote data around the peaks of the three pulses in (c).}
\end{figure}              
 
From Eq.(\ref{eta}), it is evident that if $\zeta=\frac{T_d}{T_p}$ is kept as a constant for a given $D$, then the denominator is a constant and the numerator exponentially approaches 1 when $T_p$ is approaching zero. Thus, one gains SE for a shorter $T_p$. To keep $\zeta$ constant when decreasing $T_p$, one has to increase the control intensity since $T_d=\frac{D\Gamma}{\Omega_c^2}$. Experimentally, one is limited by the available control power when choosing a shortest $T_p$, which is $\sim$200 ns in our case. Fig. 4(a) depicts the slow light transmission versus $T_p$ for an OD of 340 with $\zeta=2.3\pm 0.06$. The calculation of Eq.(\ref{eta}) with parameters \{$\zeta,\gamma_{31},\gamma_{21}$\} of \{$2.3, 0.8\Gamma,0.0002\Gamma$\} fits the data well.

Next, we address SE versus storage time. Reduction of the stray magnetic field and a small $\theta$ are two keys to prolonging the storage time\cite{Pan08}. Fig. 4(b) shows the SE versus storage time with an OD of 550 at $\theta=0.5^0$. The TBP at 50\% SE is 1200, which is the largest record to date\cite{Buchler16}. This value can be even larger if $T_p$ is shorter, which requires a higher control intensity to keep $\zeta$ fixed. In Fig. 4(c), we plot the expected TBP at 50\% SE with the results of Fig. 4(a) and a coherence time $\tau$ of 325 $\mu$s. If $T_p$ is 20 ns for an OD of 1000, the required $\Omega_c$ is 24$\Gamma$ and the TBP could be increased by a factor of 10. 

\begin{figure}
\includegraphics[width=8.5cm,viewport=0 0 520 330,clip]{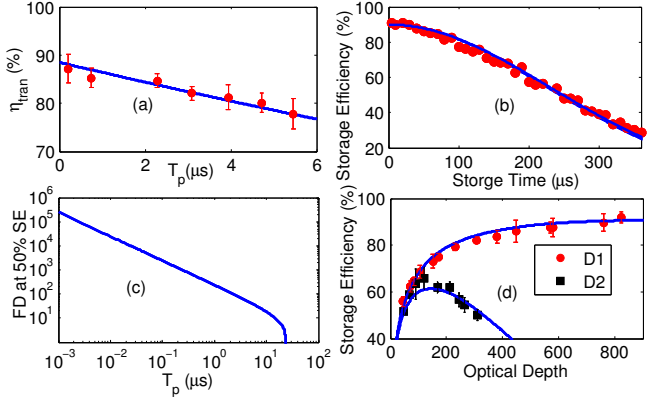}
\caption{(a) $\eta_{tran}$ versus $T_p$ for the $D_1$ scheme. The blue line is a calculation of Eq. (1) with parameters \{$\zeta,\gamma_{31},\gamma_{21}$\} of \{$2.3, 0.8\Gamma,0.0002\Gamma$\}, respectively. (b) SE versus storage time. The blue line is a fit curve of $Ae^{-t^2/\tau^2}$ with $A$=0.90 and $\tau=325\mu$s. (c) TBP at 50\% SE for the results in (a) and with $\tau=325\mu$s. (d) SE versus OD for the $D_1$ (circle) and $D_2$ (square) schemes. The line for the $D_1$ scheme is a calculation based on Eq.(1) with parameters \{$T_p,\zeta,\gamma_{21}$\} of \{207 ns, 2.7, 0.0001$\Gamma$\}, respectively. The line for the $D_2$ scheme is a calculation of Eq.(1) with parameters \{$T_p,\zeta$\} of \{207 ns, 2.7\} and $\gamma_{21}$=$\gamma_0+\frac{(48/7)\Omega_c^2}{4\delta_s^2}\gamma_{41}$, where $\gamma_0$=0.0005$\Gamma$ and $\delta_s$=2$\pi\times$251.09 MHz. The relation between $\gamma_{31}$ (or $\gamma_{41}$) and OD is mentioned in\cite{supp}.
}
\end{figure}              

Fig. 4(d) depicts the SE versus OD for the $D_1$ system (circle). We keep $\zeta$ as a constant ($\sim 2.7\pm$0.05) for all ODs by varying the control intensity. The solid line is a calculation of Eq.(1) with $\gamma_{21}=0.0001\Gamma$. From Eq.(1), it is evident that the SE approaches an asymptote value of $e^{-2\gamma_{21}T_d}$ in the high OD limit. Quantitatively, if $\gamma_{21}$ increases from $10^{-4}\Gamma$ by a factor of 10 or 100 with $T_p=200$ns, the asymptotic SE decreases from 90.7\% to 88.2\% or 66.7\%, respectively. This highlights the importance to achieve a low $\gamma_{21}$. The highest achieved SE is 92.0(1.5)\%, as shown in Fig. 4(d). SE versus OD for the $D_2$ scheme are also shown (rectangle). SE peaks at 65\% with an OD around 121 and goes down at larger ODs. The calculation based on Eq.(1) with a $\Omega_c$-dependent decoherence rate agrees well with the data\cite{supp}.

In the level scheme depicted in Fig.1(b), the off-resonant coupling of the control field on the probe transition can induce a FWM process\cite{Lett07,Novikova09}. The classical behaviors of the FWM have been well studied\cite{Lukin88,Lett07,Novikova09}. Its effect in the quantum regime has been studied theoretically\cite{Fleischhauer13} and partially tested\cite{Chen14}. Noise photon due to FWM, which exponentially grows with the parameter $x=D\Gamma/\delta_{hf}$, is detrimental to quantum memories, where $\delta_{hf}$ is the hyperfine splitting of the ground states. To reduce the FWM, it is better to choose a system which has a larger $\delta_{hf}/\Gamma$\cite{Fleischhauer13}. Among the stable alkali, cesium has the most favorable value.

\begin{figure}
\includegraphics[width=8cm,viewport=230 5 763 370,clip]{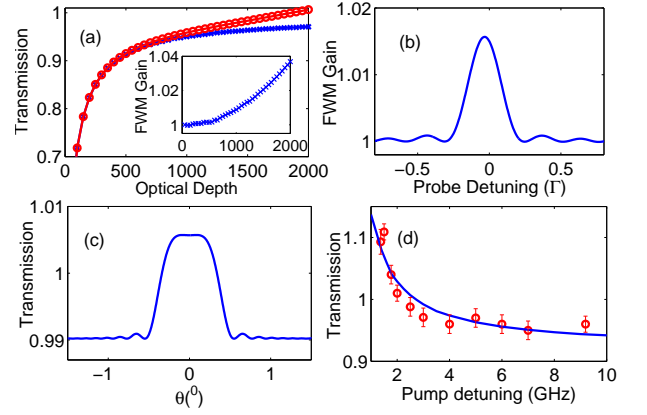}
\caption
{(a) Numerical calculations of $\eta_{tran}$ versus OD with(red circle) and without(blue cross) the presence of FWM. Perfect phase matching is assumed and $T_p$ is 200 ns. $\Omega_c$ is adjusted for each OD to keep $\zeta$=2.7. Inset shows the ratio of those two traces, reflecting the FWM gain versus OD. The results in (b)-(c) are based on the steady-state calculation including the phase miss-matching. The parameters \{$D,\Omega_c,\gamma_{21}$\} are \{1000,6.72$\Gamma$,0.0002$\Gamma$\}, respectively. (b) FWM gain versus $\delta_p$ at $\theta=0^0$. (c) Probe transmission versus $\theta$ for $\delta_p$= 0.04$\Gamma$. (d) Peak probe transmission versus the pump detuning for an OD of 603. The pump power is half of the control power. The blue solid curve is a numerical calculation with $D=600, \theta=0.5^0, \Omega_c=6.2\Gamma$, and $\gamma_{21}=0.0005 \Gamma$.   
}
\end{figure}  

To evaluate the FWM effect, we first perform a semi-classical calculation of the probe gain\cite{Novikova09,supp,Lukin88,Boyer13}. A larger gain indicates a more reduction of the fidelity\cite{Fleischhauer13}. Fig. 5(a) depicts a numerical calculation of the probe transmission versus OD for $T_p$=200 ns and $\zeta$=2.7 under perfect phase matching condition. The red and blue traces indicate the transmission with and without the presence of FWM. Inset shows the ratio of the two traces, indicating the FWM gain versus OD. Even at an OD of 1000, the gain is only up to 0.9\%. We also perform a steady-state calculation of the FWM gain including the phase miss-matching\cite{Boyer13,supp}. Fig. 5(b) depicts the FWM gain versus the probe detuning for $D$=1000, $\Omega_c=6.72\Gamma$, $\gamma_{21}=0.0002\Gamma$ and $\theta=0^0$, where $\theta$ is the angle between control and probe beams. The maximum FWM gain is $\sim$1.5\%, occurred at $\delta_p=0.04\Gamma$. Fig. 5(c) depicts the probe transmission versus $\theta$ for $\delta_p= 0.04\Gamma$. For $\theta \geq 0.5^0$, the FWM can be greatly suppressed because the phase matching condition is not satisfied. This provides an useful way to reduce the complications of FWM in quantum memory applications.   

The FWM gain of $\sim$(1-2)\% is within the uncertainty of the transmission determination and is thus difficult to be identified experimentally. Therefore, we design an experiment to allow a quantitative evaluation of the FWM gain. We add another pump beam overlapping with the control beam. Its detuning relative to the probe transition can be tuned such that we could obtain a measurable FWM gain at a smaller detuning and study how it varies with the detuning. At a detuning of 9.192 GHz, the role of the pump is the same as the control beam. Thus, the FWM gain introduced by the control field can be estimated experimentally. Fig.5(d) depicts such a measurement with an OD of $\sim$600 and a pump power equal to half of the control power. The blue curve is a corresponding numerical calculation with $\theta=0.5^0, \Omega_c=6.2\Gamma$. At a detuning of 9.192 GHz, the FWM gain due to the pump is $\lesssim 0.6\%$. Scaled to the case with an OD of 816 for a SE of 92\%, the estimated FWM gain is $< 3\%$. This provides an experimental support that FWM gain is negligible in our experiment. More details are described in\cite{supp}.  

In summary, we achieve a high-efficiency 92(1.5)\% EIT-based optical memory with an almost negligible FWM effect ($< 3\%$). We expect this result could be extended to regime of quantum storage\cite{Lvovsky08,Lvovsky09,Pan11,Du11,Du12,Laurat18}. Together with long storage times\cite{Kuzmich13}, EIT-based quantum memory could potentially realize some important applications\cite{Kimble08,Tittel13,DLCZ,Gisin11,Nunn13} in quantum information science.            

This work was supported by MOST of Taiwan under grant numbers 103-2119-M-001-012 and 103-2112-M-001-010-MY3. This work was done under a collaboration project (Science Vanguard Research Program of MOST) with I.A.Y. as the project leader and Y-C.C. and Y-F.C. as the subproject leaders. Correspondence of the project contents can be addressed to I.A.Y. Correspondence and requests for materials of this work can be addressed to Y-C.C. We also acknowledge the support from NCTS.

\end{document}


\title{Highly Efficient Coherent Optical Memory Based on Electromagnetically Induced Transparency: Supplementary Material}

\maketitle
\setcounter{equation}{0}
\setcounter{figure}{0}
\setcounter{table}{0}
\setcounter{page}{1}
\makeatletter
\renewcommand{\theequation}{S\arabic{equation}}
\renewcommand{\thefigure}{S\arabic{figure}}
\renewcommand{\bibnumfmt}[1]{[S#1]}
\renewcommand{\citenumfont}[1]{S#1}
\subsection{Slow light transmission in a three-level $\Lambda$-type system}
In a $\Lambda$-type three-level system, the classical probe field (with frequency $\omega_p$ and Rabi frequency $\Omega_p$) drives the ground state $|1\rangle$ to the excited state $|3\rangle$, and the classical control field (with frequency $\omega_c$ and Rabi frequency $\Omega_c$) drives another ground state $|2\rangle$ to  $|3\rangle$, where the Rabi frequencies $\Omega_{p(c)}=-\vec{d}_{13(23)}\cdot\vec{E}_{p(c)}/\hbar$ are assumed to be real. The transition frequencies from state $|1\rangle$ and $|2\rangle$ to $|3\rangle$ are denoted as $\omega_{31}$ and $\omega_{32}$, respectively. Under the rotating-wave approximation and in the interaction picture, the Hamiltonian can be expressed as 
\begin{equation}\label{Hamiltonian} 
\begin{bmatrix}
       0 & 0 & -\frac{\hbar\Omega_p}{2}           \\[0.3em]
       0 & -\hbar\delta_2           & -\frac{\hbar\Omega_c}{2} \\[0.3em]
        -\frac{\hbar\Omega_p}{2}           &  -\frac{\hbar\Omega_c}{2} & -\hbar\delta_p
     \end{bmatrix},
\end{equation} 
where $\delta_{p(c)}=\omega_p(c)-\omega_{31(32)}$ is the probe (control) detuning and $\delta_2=\delta_p-\delta_c$ is the two-photon detuning. The relaxation terms for the density matrix are denoted as,
\begin{equation} 
\begin{bmatrix}\label{relaxation}
       \Gamma_{31}\sigma_{33} & -\gamma_{12}\sigma_{12} & -\gamma_{13}\sigma_{13}           \\[0.3em]
       -\gamma_{21}\sigma_{21} & \Gamma_{32}\sigma_{33}           & -\gamma_{23}\sigma_{23} \\[0.3em]
        -\gamma_{31}\sigma_{31}           &  -\gamma_{32}\sigma_{32} & -\Gamma\sigma_{33}
     \end{bmatrix},
\end{equation}  
with $\Gamma=\Gamma_{31}+\Gamma_{32}$. Under the weak probe assumption ($\Omega_p<<\Omega_c$), the relevant first-order optical Bloch equations (OBEs) are,
\begin{equation}\label{rho31td}
\frac{d\sigma_{31}}{dt}=(i\delta_{p}-\gamma_{31})\sigma_{31}+\frac{i}{2}\Omega_c\sigma_{21}+\frac{i}{2}\Omega_p,
\end{equation}
\begin{equation}\label{rho21td}
\frac{d\sigma_{21}}{dt}=(i\delta_{2}-\gamma_{21})\sigma_{21}+\frac{i}{2}\Omega_c\sigma_{31}.
\end{equation}
Under the slowly-varying envelope approximation, the Maxwell equation for the probe field is,
\begin{equation}\label{Maxwelltd}
\frac{\partial \Omega_p}{\partial z}+\frac{1}{c}\frac{\partial \Omega_p}{\partial t}=i\frac{D\Gamma}{2L}\sigma_{31},
\end{equation}
where $D$ is the optical depth of the atomic media and $L$ is the media length. Since $\Omega_p <<\Omega_c$, one can treat $\Omega_c$ as a constant. Taking the Fourier transform on the two atomic coherences ($\sigma_{31}$ and $\sigma_{21}$) and the probe Rabi frequency $\Omega_p$ to frequency domain, e.g. $R_{31}=1/\sqrt{2\pi}\int_{-\infty}^{\infty}\sigma_{31}e^{i\omega t}dt$, the OBEs and Maxwell equation read as follows:
\begin{equation}\label{rho31fd}
-i\omega R_{31}=(i\delta_p-\gamma_{31})R_{31}+\frac{i}{2}\Omega_{c}R_{21}+\frac{i}{2}W_p,
\end{equation}
\begin{equation}\label{rho21fd}
-i\omega R_{21}=(i\delta_2-\gamma_{21})R_{21}+\frac{i}{2}\Omega_{c}R_{31},
\end{equation}
\begin{equation}\label{Maxwellfd}
\frac{\partial W_p}{\partial z}-\frac{i\omega}{c}W_p=i\frac{D\Gamma}{2L}R_{31}.
\end{equation}
By solving Eqs.(\ref{rho31fd}) and (\ref{rho21fd}), one obtains the expression for $R_{31}$,
\begin{equation}\label{R31}
R_{31}(\omega,z)=\frac{-[i(\omega+\delta_2)-\gamma_{21}]iW_p(\omega,z)/2}{[i(\omega+\delta_p)-\gamma_{31}][i(\omega+\delta_2)-\gamma_{21}]+\Omega_c^2/4}.
\end{equation}
Putting this into Eq.(\ref{Maxwellfd}) and integrating over $z$, one obtains the solution of $W_p(\omega,z)$,
\begin{equation}\label{Wpfd}
W_p(\omega,z)=W_p(\omega,0)exp[{\frac{i\omega z}{c}+\frac{Dz\Gamma}{4L}\frac{i(\omega+\delta_2)-\gamma_{21}}{[i(\omega+\delta_p)-\gamma_{31}][i(\omega+\delta_2)-\gamma_{21}]+\Omega_c^2/4}}].
\end{equation}
The steady-state EIT transmission spectrum can be obtained by setting $\omega=0$ and $z=L$ in Eq.(\ref{Wpfd}), which is
\begin{equation}\label{PowerSpec}
T(\delta_p)=Exp\{\frac{D\Gamma}{2} Re(\frac{i(\delta_p-\delta_c)-\gamma_{21}}{(i\delta_p-\gamma_{31})(i(\delta_p-\delta_c)-\gamma_{21})+\frac{\Omega_c^2}{4}})\},
\end{equation}
where $Re()$ stands for the real part of the expression inside the bracket.
From Eq.(\ref{PowerSpec}) with $\delta_c=0$, one can show that the FWHM frequency width of the EIT transmission spectrum is,
\begin{equation}\label{EITbw}
\Delta\omega_{EIT}\cong\sqrt{\frac{ln2}{2}}\frac{\Omega_c^2}{\sqrt{D\gamma_{31}\Gamma}},
\end{equation}
where we assume $\Omega_c>>\Delta\omega_{EIT}$ and $\Omega_c>>4\gamma_{31}\gamma_{21}$.
We assume the input probe pulse is a Gaussian waveform with an intensity FWHM duration of $T_p$, i.e.
\begin{equation}\label{gaussiantd}
\Omega_p(t,z=0)=\Omega_{p0}exp(-2ln2\frac{t^2}{T_p^2}).
\end{equation}
The Fourier transform of the input probe pulse can be calculated to be,
\begin{equation}\label{Wpinifd}
W_p(\omega,z=0)=\frac{\Omega_{p0}T_p}{\sqrt{4ln2}}exp(-\frac{\omega^2T_p^2}{8ln2}).
\end{equation}
Putting Eq.(\ref{Wpinifd}) into Eq.(\ref{Wpfd}) and taking the inverse Fourier transform, one obtains the solution of the probe pulse after passing through an EIT media as, 
\begin{equation}\label{Opsol}
\Omega_p(t,z=L)=\frac{1}{\sqrt{2\pi}}\frac{\Omega_{p0}T_p}{\sqrt{4ln2}}\int_{-\infty}^{\infty}d\omega exp\{-i\omega t-\frac{\omega^2T_p^2}{8ln2}+\frac{i\omega L}{c}+\frac{(i(\omega+\delta_2)-\gamma_{21})D\Gamma}{4(i(\omega+\delta_p)-\gamma_{31})(i(\omega+\delta_2)-\gamma_{21})+\Omega_c^2}\}.
\end{equation} 
The output probe pulse with arbitrary parameters can be numerically calculated by this relation. Under the special case with $\delta_p=0=\delta_c$, one can expand the EIT media response function $f(\omega)=i\frac{D\Gamma}{2}\frac{R_{31}}{W_p}$ with respect to $\omega$ as follows,
\begin{equation}\label{fw}
f(\omega)=\frac{(i\omega-\gamma_{21})D\Gamma}{4(i\omega-\gamma_{31})(i\omega-\gamma_{21})+\Omega_c^2},
\end{equation}
\begin{equation}           
=\frac{-\gamma_{21}D\Gamma}{\Omega_c^2+4\gamma_{21}\gamma_{31}}+i\frac{D\Gamma(\Omega_c^2-4\gamma_{21}^2)}{(\Omega_c^2+4\gamma_{21}\gamma_{31})^2}\omega-\frac{4D\Gamma(\gamma_{31}\Omega_c^2+2\gamma_{31}^2\gamma_{21}-4\gamma_{21}^3)}{(\Omega_c^2+4\gamma_{21}\gamma_{31})^3}\omega^2+O(\omega^3)\\ 
\end{equation}
\begin{equation}\label{fw2}
\cong -\frac{\gamma_{21}D\Gamma}{\Omega_c^2}+i\frac{D\Gamma}{\Omega_c^2}\omega-\frac{4D\gamma_{31}\Gamma}{\Omega_c^4}\omega^2+O(\omega^3),
\end{equation}
where it is assumed that $\Omega_c^2>>4\gamma_{21}\gamma_{31}$. If we keep the dispersion relation up to the $\omega^2$ term, Eq.(\ref{Opsol}) can be analytically integrated to become,
\begin{equation}\label{Opfinal}
\Omega_p(t,z=L)=\frac{\Omega_{p0}}{\beta}exp(-\gamma_{21}\frac{D\Gamma}{\Omega_c^2})exp[-2ln2(\frac{t-T_d}{\beta T_p})^2],
\end{equation}
where we have defined
\begin{equation}\label{beta}
\beta=\sqrt{1+\frac{32ln2D\Gamma\gamma_{31}}{T_p^2\Omega_c^4}}
\end{equation}
\begin{equation}\label{Td}
T_d=\frac{L}{v_g}=\frac{L}{c}+\frac{D\Gamma}{\Omega_c^2}.
\end{equation}
From Eq.(\ref{Opfinal}), it is evident to see that the amplitude of the slow light pulse decreases by a factor of $\beta$ times a factor due to the finite ground-state decoherence rate and its duration broadens by a factor of $\beta$. 
If the group velocity $v_g<<c$, then the group delay $T_d\cong\frac{D\Gamma}{\Omega_c^2}$. Integrating Eq.(\ref{Opfinal}) over the time, one can obtain the slow light energy transmission as follows:
\begin{equation}
T=\frac{exp(-2\gamma_{21}T_d)}{\beta}=\frac{exp(-2\gamma_{21}T_d)}{\sqrt{1+32ln2\frac{\gamma_{31}}{\Gamma}\frac{\zeta^2}{D}}},
\end{equation}
where $\zeta\equiv T_d/T_p$. Combining this relation with Eq.(\ref{EITbw}), one obtains,
\begin{equation}\label{T2}
T=\frac{exp(-2\gamma_{21}T_d)}{\sqrt{1+(\frac{4ln2}{T_p\Delta\omega_{EIT}})^2}}.
\end{equation}
It is evident to see that the finite ground-state decoherence rate and the finite EIT transparent bandwidth are the two limiting factors for the slow light transmission.

\begin{figure}
\includegraphics[width=8.5cm,viewport=300 100 720 330,clip]{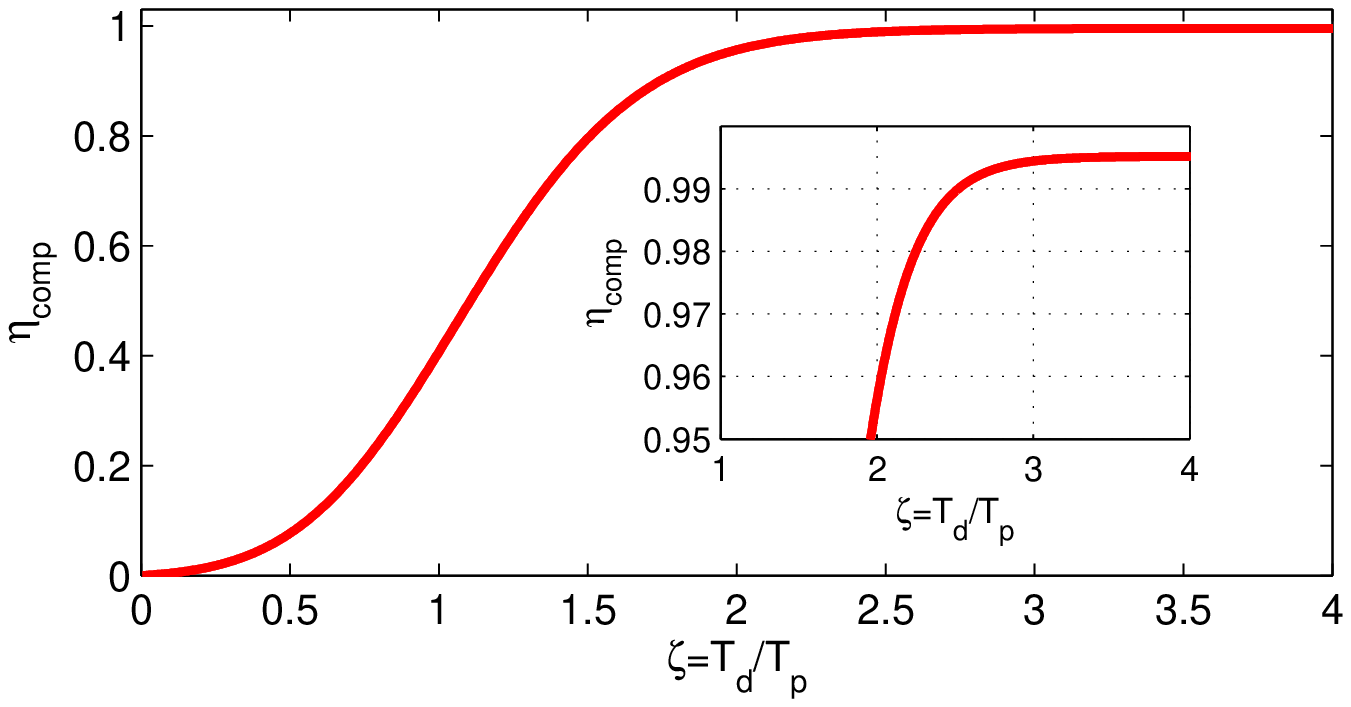}
\caption{$\eta_{comp}$ versus $\zeta$ for the case with $\kappa=1.1$ and $D$=100.}
\end{figure}

\subsection{Efficiency due to cutoff of the pulse edges during storage}
At time $t=0$, the peak of a Gaussian probe pulse enters the media. At time $t=t_c$, one turns off the control field and stores the major part of the pulse inside the media. A small portion of the front edge of the pulse has passed through the media, while a small portion of the rear edge has not yet arrived at the media. These two parts of pulse cannot be stored into the media. The fractional energy being stored can be written as follows:

\begin{equation}
\eta_{comp}=1-\frac{\int_{-\infty}^{t_c}\Omega_p^{2}(z=0,t)dt}{\int_{-\infty}^{\infty}\Omega_p^2(z=0,t)dt}-\frac{\int_{t_c}^{\infty}\Omega_p^{2}(z=L,t)dt}{\int_{-\infty}^{\infty}\Omega_p^2(z=L,t)dt}.
\end{equation}
Using Eqs. (\ref{gaussiantd}) and (\ref{Opfinal}) and the definition of the error function, one obtains
\begin{equation}
\eta_{comp}=\frac{1}{2}[erf(2\sqrt{ln2}\kappa)+erf(2\sqrt{ln2}\frac{\zeta-\kappa}{\beta})],
\end{equation} 
where $\kappa=\frac{t_c}{T_p}$, $\zeta=\frac{T_d}{T_p}$. With a large enough OD, it is possible to find that with $\kappa$ and $\zeta$ larger than a critical value then $\eta_{comp}$ can be larger than 0.99. With a larger OD, the critical value for $\zeta$ at a given $\kappa$ can be smaller. Fig. S2 shows an example for $\eta_{comp}$ versus $\zeta$ for $\kappa$=1.1 and OD=100. It can be seen that with a  $\zeta\gtrsim$2.5, more than 99\% of the pulse energy could be stored into the medium.  

\subsection{Experimental setup and timing sequence}
Our experiment is based on a vapor-cell two-dimensional magneto-optical trap (MOT) of cesium\cite{Yu08}. The total powers of the trapping and repumping beams after the single mode fiber are $\sim$ 350 and 50 mW, respectively. The diameters for both beams are $\sim$22 mm. We typically trap $\sim 5\times 10^9$ atoms with a cigar-shaped cloud of dimensions $\sim 3\times 3\times 14$ mm. The temperature of the atomic clouds is typically $\sim$150 $\mu$K, measured by the time of flight method with absorption imgaging. To increase the optical depth, we have utilized temporally dark and magnetically compressed MOT techniques, and both hyperfine-state and Zeeman-state optical pumping to prepare most population in the Zeeman state $|F=3, m=3\rangle$\cite{Chen14b}. The reason for preparation of population in such a state is described in the following.

In the $D_1$-line EIT system (see Fig. 1(b) of the main text), if one prepares the population in the $|F=3, m=3\rangle$ ground state and chooses the same $\sigma^{+}$-polarized light for both the control and probe beams, then the EIT system only involves three states ($|F=3, m=3\rangle$, $|F=4, m=3\rangle$ and $|F'=4, m=4\rangle$) and the control beam is completely free from the so-called photon switching effect\cite{Harris98} since there is no any other nearby excited state the control beam can off-resonantly couple to. It is thus free from the degradation of storage efficiency due to the control-power-dependent decoherence rate. It should be noted that the probe drives the $|F=3, m=3\rangle\rightarrow|F'=4, m=4\rangle$ transition, which has the largest Clebsch-Gordan coefficient in the $D_1$ transition. Such a choice allows a largest OD for the probe transition. If atoms are prepared in the $|F=4, m=4\rangle$ ground state, the involved three levels for the EIT system could be $|F=3, m=2\rangle$, $|F=4, m=4\rangle$ and $|F'=3, m=3\rangle$ with probe driving the $|F=4, m=4\rangle\rightarrow|F'=3, m=3\rangle$ transition, since  it has the same largest Clebsch-Gordan coefficient as the $|F=3, m=3\rangle\rightarrow|F'=4, m=4\rangle$ transition. However, the control field can off-resonantly couple to $|F=3, m=2\rangle\rightarrow|F'=4, m=3\rangle$ transition and induce the control-power-dependent decoherence rate. Therefore, preparing population in the $|F=3, m=3\rangle$ state and choosing the energy-level scheme as in Fig. 1(b) is the best choice for implementing an EIT system in terms of maximizing the probe OD and minimizing the $\Omega_c$-dependent decoherence rate. 

To prepare most of the population in the F=3 ground state, the repumping beams are turned off 600 $\mu$s before turning off the trapping beams. The EIT control beam are turned on after the repumping beams are turned off. One additional depumping beam, which drives the $|F=4\rangle\rightarrow|F'=4\rangle$ transition of the $D_2$ line, is also turned on at the same timing as the control beam. The depumping beam has a power of $\sim$7 mW and a diameter of $\sim$ 10mm to cover the whole atomic cloud. It is coupled into the horizontal trapping beams through one polarizing beam splitter. The depumping beam is used to help the hyperfine-state optical pumping to the F=3 ground state. The depumping beam is off during the slow light experiment. To diagnose the efficiency of preparing population into the F=3 ground state, we apply one probe pulse which drives the $|F=4\rangle\rightarrow|F'=4\rangle$ transition of the $D_2$ line. If there is still some fractions of population remaining in the F=4 ground state after the hyperfine pumping stage, part of the front edge of this probe pulse should be absorbed due to the optical pumping effect\cite{Chen01}. The absorbed area is proportional to the population in the F=4 ground state. As a comparison, we also apply the same probe pulse but without hyperfine pumping at all such that almost all population are in the F=4 ground state. From the ratio of the two absorbed area, we can determine the efficiency of the hyperfine optical pumping. At an optical depth around 816, we found that more than 99\% of the population are being pumped to the F=3 ground state.

The Zeeman optical pumping beam drives the $|F=3\rangle \rightarrow |F'=2\rangle$ transition of the $D_2$ line. The power of the Zeeman pumping beam is up to 45 mW and its diameter is 5 mm. It is nearly circularly polarized. It intersects with the EIT probe beam by about $\sim 4^0$ to induce both $\sigma^+$ and $\pi$ transitions with the Zeeman state $|F=3, m=3\rangle$ being the only dark state\cite{Chu98}. It is typically turned on for 20 $\mu$s before the slow light experiment. We have observed a gradual degradation of the effectiveness of the Zeeman optical pumping as OD increases, possibly due to the effect of radiation trapping. We have performed microwave spectroscopy to diagnose the population distribution among the Zeeman sublevels after the optical pumping. A magnetic field of 360 mG is applied along the propagation direction of the probe beam to split the Zeeman sublevels. The microwave signal is amplified to 2.5 W and sent through a horn antenna. After Zeeman optical pumping, a microwave pulse with a duration of 70 $\mu$s is applied to pump the population in the F=3 ground hyperfine state to the F=4 ground state. The frequency of the MOT laser is tuned to resonance and is turned on for 50 $\mu$s. The fluorescence signal is collected and measured by a CCD camera. By scanning the microwave frequency through 9.192 GHz, the whole spectrum with 15 peaks is obtained. Considering the oscillator strength of different microwave transition\cite{Roberts49}, the Zeeman population can be determined. Fig. S1(a) and (b) shows the spectrum without and with Zeeman optical pumping, respectively with an OD of 580. In this case, we estimate that the population in the $|F=3, m=3\rangle$ state is $\sim (74\pm3)\%$. For the maximum OD of 816 in Fig.4(d) of the main text, the population in the $|F=3,m=3\rangle$ state is $\sim (67\pm4)\%$.   

\begin{figure}
\includegraphics[width=12cm,viewport=220 40 840 380,clip]{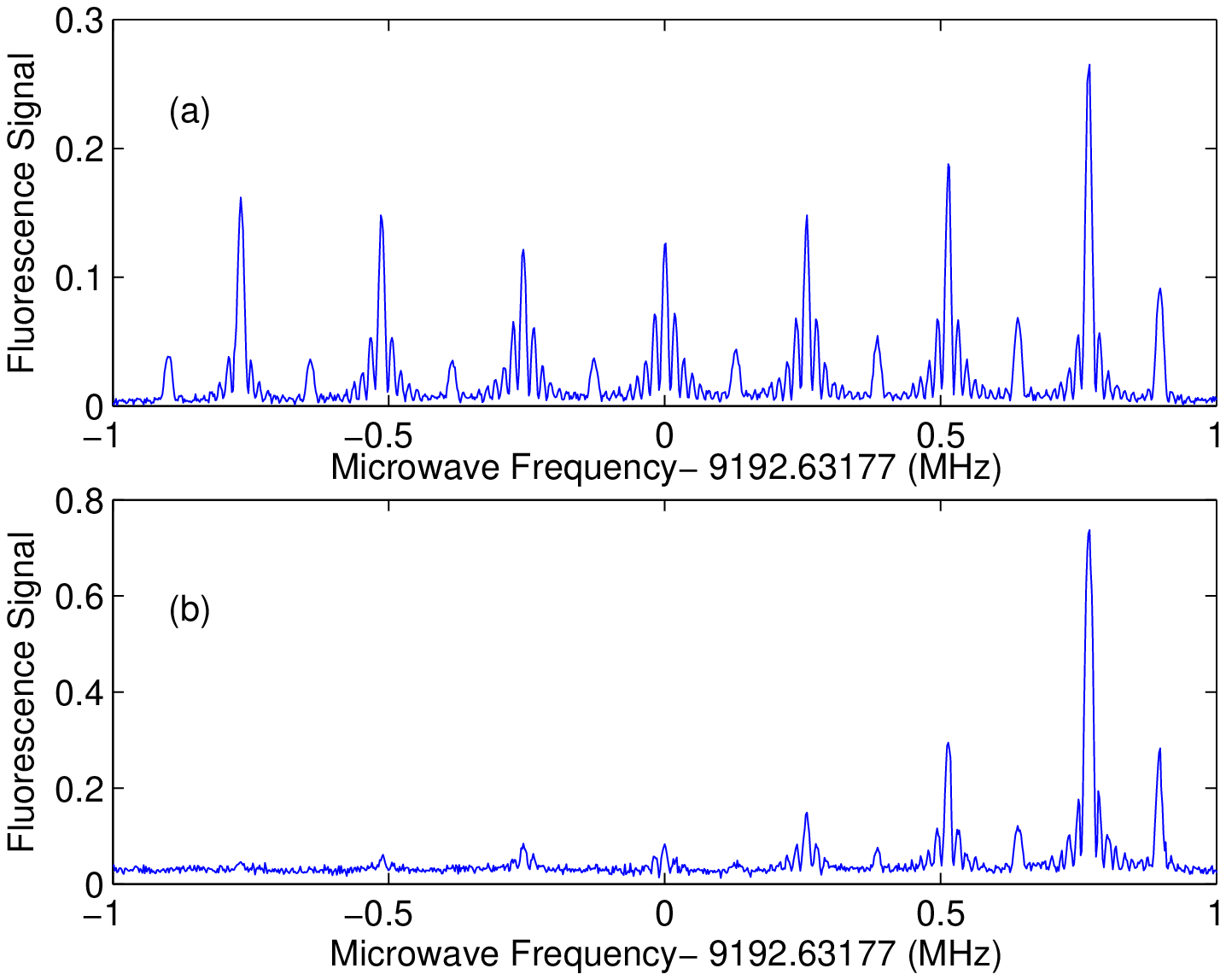}
\caption{(a) and (b): Representative microwave spectrum without and with Zeeman optical pumping, respectively. The optical depth is 580(45) in this case. The measured population in the $F=3,m=3\rangle$ is $\sim (74\pm3)\%$ for (b).}
\end{figure}
     
Two master lasers are locked to cesium saturation absorption spectrometer at $D_1$ and $D_2$ line, respectively. The $D_1$ and $D_2$ control laser are injection locked by the corresponding master laser. Part of each master light passes through a fiber electro-optic modulator (EOM) operated around 9 GHz and its $+1$ sideband injection-lock the $D_1$ and $D_2$ probe laser, respectively. 
It is important to injection lock the control laser with the light directly from the master laser, instead of using the sideband after passing through the fiber EOM. For the later arrangement, the control laser may contain a small fraction in energy at the same frequency as the carrier frequency. Because the control power is relatively strong and the frequency of the unwanted spectral component is close to the probe transition, even a small fraction of contamination in the control laser may introduce a four-wave mixing gain in the probe and lead to an incorrect determination of the storage efficiency. Some acousto-optic modulators (AOMs) are used for power switching and frequency shifting such that the frequencies of the control beams are on the $|F=4\rangle\rightarrow|F'=4\rangle$ transition and that of the probe beams are on the $|F=3\rangle\rightarrow|F'=4\rangle$ transition of the $D_1$ and $D_2$ line, respectively. With the injection locking technique and good frequency stabilities for the AOM and EOM drivers, the FWHM linewidth of the beatnote between the control and probe lasers is measured to be 10 Hz, limited by the resolution bandwidth of the spectrum analyzer. The good mutual coherence between the control and probe lasers is one of the key to obtain a small ground-state decoherence rate.  

The linewidth of the master laser also plays an important role in obtaining a high SE. During the early operations of the experiment, we used a 895 nm homemade external cavity diode laser (ECDL) as the $D_1$-line master laser. The laser linewidths were observed to drift within the range of 200 kHz to 3 MHz in a few minutes time scale. The linewidth is measured by the delayed self-heterodyne method\cite{Richter86}. The long-term laser frequency fluctuations may also contribute to the overall laser linewidth since the data averaging takes about 34 s. From the frequency measurement of the probe laser by a wavelength meter (HighFinesse WS7), we estimate that the laser central frequency fluctuation is $\sim$1 MHz. Both the short-term linewidth and long-term fluctuations contribute to the overall laser linewidth. At an OD of $\sim$ 300, the SE can vary up to 6$\%$ within that observed linewidth range. The variation in SE is consistent with Eq. (1) if one includes the laser linewidth ($\Gamma_{L}$) into the decoherence rate $\gamma_{31}=\frac{\Gamma+\Gamma_{L}}{2}$. During the later operations, we have replaced the 895 nm master laser by a commercial one (Toptica DL pro), whose linewidth is checked to be less 100 kHz during the whole operation. Our 852-nm master laser is a home-made ECDL. Its linewidth is measured to be less than 1 MHz. 
    
The experiment runs at a periodic manner with the slow and stored light measurements taken at 1.5 ms after the quadrupole magnetic field
of the MOT has been turned off to reduce the ground-state decoherence rate due to the inhomogeneity of the magnetic field. Some electronics are used to reduce the $e^{-1}$ turn-off time to 200 $\mu$s. We avoid to put metallic components near the cell region to minimize the induced eddy currents. Three pairs of magnetic compensation coils are used to minimize the stray magnetic field. Optimization of the stray magnetic field compensation is performed by iteratively fine tuning of the currents through the compensation coils and prolonging the storage time. To reduce the ac magnetic noises due to the 60 Hz power line, the measurement is synchronized to it and is run at a 7.5 Hz repetition rate. 

When taking the EIT spectrum, the probe pulses of square waveform with a 100-$\mu$s duration are applied. The probe power at 35-40 $\mu$s after being turned on are measured to determine the probe transmission in order to obtain the steady-state response. By varying the probe frequency through a double-passed AOM and repeating the measurement, the EIT spectrum is obtained.

We also utilize the beatnote interferometer\cite{Yu05,Yu06} to check the coherence property of the slow and stored-and-retrieved pulses. In the beatnote measurement, a continuous reference beam, red-detuned by 200 MHz relative to the probe transition, is spatially combined with the probe pulse at one beam splitter and sent through the atomic cloud (Fig.1 (c) of the main text). The light is directed into an avalanche photodetector (Hamamastu C5658, 1 GHz bandwidth) without passing through the etalon filter. The other part of light at the beam splitter is detected by another photodetector (NewFocus 1801, 125 MHz bandwidth). This beat signal is served as a phase reference and is used to trigger the oscilloscope. The beatnote signal of Fig. 3 (c) of the main text is measured this way.     

\subsection{Determination of the experimental parameters}
The experimental parameters are determined by the joint fitting of EIT spectra and slow light traces. By assuming a different value of $\gamma_{31}$ and fitting the EIT spectrum to Eq. (S11), one observes that the parameters that could be uniquely determined by the spectral fitting are $\Omega_c$, $\delta_c$, $D\gamma_{21}$, and $D\gamma_{31}$. With the obtained $\Omega_c$, the group delay Eq.(S21), the pulse broadening factor Eq.(S20), and the transmission efficiency of the slow light Eq.(S22), one can completely determine the parameters $D$, $\gamma_{21}$, and $\gamma_{31}$. Once these three parameters are determined by the slow light properties, one can check the consistencies with the $D\gamma_{31}$ determined by the spectral fitting. The observed discrepancies of $D\gamma_{31}$ are typically within $\pm$15\%, which are acceptable. The parameters $D\gamma_{21}$ determined by EIT spectral fitting are not reliable, especially for the data of $D_1$ system, since the EIT peak transmission are all around unity. For the parameter $\gamma_{21}$, we rely on that determined by the slow light trace. The data shown in Fig. 4(d) of the main text are determined in this way. By performing a polynomial fit to power of two for the determined $\gamma_{31}$ as a function of OD, the relation is $\gamma_{31}=(0.70+4.20\times 10^{-5}D+4.87\times 10^{-7}D^2)\Gamma$ for the $D_1$ system and $\gamma_{31}$=$\gamma_{41}$=$(0.70+3.90\times 10^{-4}D+1.47\times 10^{-6}D^2)\Gamma$ for the $D_2$ system. We assume $\gamma_{41}$ is the same as $\gamma_{31}$ in the $D_2$ system. We notice that $\gamma_{31}$ increases as OD increases. We speculate that this spectral broadening may due to the cooperative effect by the resonant dipole-dipole interactions\cite{Jennewein16}. Our theoretical model does not include such an interaction but its effect is effectively incorporated through the variation in $\gamma_{31}$ with OD. The role of cooperative effect in EIT certainly deserves a further study but is not the focus of this work.  
For smaller ODs, the determined $\gamma_{31}$ approaches to $\sim 0.7\Gamma$. The reason may due to the overall laser linewidth including the short-term laser linewidth and laser frequency fluctuations during the locked condition. As an independent check, we also take the two-level absorption spectrum of the probe. The OD is kept small ($\sim$1) such that the maximum absorption is not complete zero ($\sim$0.3-0.4). Thus, the spectral linewidth$\gamma_{31}$ can be directly fitted from the spectra without ambiguity. The obtained $\gamma_{31}$ is 0.74$\pm 0.03 \Gamma$, consistent with what mentioned above.

\subsection{Filtering optics of the control beam} 
We have kept the angle $\theta=0.5^0$ between the control and probe beams to reduce the decoherence effect due to the atomic motions. With such a small angle, the leakage of the strong control beam into the PMT for probe detection become an issue. Many arrangements are used to minimize this leakage. First, the control beam is kept well collimated and the probe beam is focused around the atomic cloud. The $e^{-2}$ diameter for the control beam is 1.1 mm and the $e^{-2}$ diameter of the probe beam is $\sim$100 $\mu$m. After passing through the cell, there is a lens to make the probe beam collimated but to focus the control beam. Around the focal point of the control beam, a window with a small black dot of diameter $300\mu$m is used to reduce the control beam leakage by $\sim$38 dB. The probe beam propagates around 6.9 m and pass through two iris before it arrives at a temperature-stabilized etalon filter(Quantaser FPE001). The etalon reduces the control leakage by 35 dB and its maximum transmission is 58\%. The probe output after the etalon is coupled into a multimode fiber and detected by a PMT (Hamamatsu R636-10). The probe transmission efficiency of the cell window, window with a black dot, and fiber coupling are 95\%, 82\%, and 80\%, respectively. The probe beam propagates in ambient environment for $\sim$7 m and we have observed $\sim 13\%$ loss due to the possible water absorption loss around 894.6 nm\cite{Schermaul01}. The overall collection efficiency of the probe beam after these filtering components is 31\%.   

 \subsection{Slow light transmission in a N-type four-level system}
Considering the off-resonant excitation of the control field from the state $|2\rangle$ to an additional excited state $|4\rangle$ as that shown in Fig.1 (a), the level scheme is a $N$-type four-level system in which the control field also acting as the role of the switching field which off-resonantly drives $|2\rangle\rightarrow|4\rangle$ transition\cite{Harris98}. In the $N$-type system, the probe field has one additional loss channel due to the multi-photon process from state $|1\rangle\rightarrow|3\rangle\rightarrow|2\rangle\rightarrow|4\rangle$ and then spontaneous decay. The Rabi frequency of the switching field is $\Omega_s=\epsilon\Omega_c$, where $\epsilon$ is the ratio of the Clebsch-Gordan coefficient of the switching transition to that of the control transition. In the level scheme of Fig.1(a), $\epsilon$ is $\sqrt{48/7}$.

Similar to the procedures before, one can write down the frequency-domain first-order perturbative OBEs as,
\begin{equation}\label{r31N}
-i\omega R_{31}=(i\delta_p-\gamma_{31})R_{31}+\frac{i\Omega_c}{2}R_{21}+\frac{i}{2}W_p,
\end{equation}
\begin{equation}\label{r21N}
-i\omega R_{21}=(i\delta_2-\gamma_{21})R_{21}+\frac{i\Omega_c}{2}R_{31}+\frac{i\Omega_s}{2}R_{41},
\end{equation}
\begin{equation}\label{r41N}
-i\omega R_{41}=(i\delta_3-\gamma_{41})R_{41}+\frac{i\Omega_s}{2}R_{21},
\end{equation}
where the three-photon detuning $\delta_3=\delta_p-\delta_c+\delta_s$ and $\delta_s=\omega_s-\omega_{42}$. By solving Eqs.(\ref{r31N})-(\ref{r41N}), one obtains,

\begin{equation}\label{R31N}
R_{31}=-\frac{iW_p}{2}\frac{[i(\omega+\delta_2)-\gamma_{21}][i(\omega+\delta_3)-\gamma_{41}]+\Omega_s^{2}/4}{[i(\omega+\delta_p)-\gamma_{31}]\{{[i(\omega+\delta_2)-\gamma_{21}][i(\omega+\delta_3)-\gamma_{41}]+\Omega_s^{2}/4\}+[i(\omega+\delta_3)-\gamma_{41}]\Omega_c^{2}/4}}
\end{equation}

Putting it into Eq. (\ref{Maxwellfd}) and setting $\omega=0$, one obtains the steady-state spectrum for the N-type four-level system as, 
\begin{equation}\label{Nspectrum}
T(\delta_p)=Exp\{\frac{D\Gamma}{2}Re(\frac{(i\delta_2-\gamma_{21})(i\delta_3-\gamma_{41})+\Omega_s^{2}/4}{(i\delta_p-\gamma_{31})[{(i\delta_2-\gamma_{21})(i\delta_3-\gamma_{41})+\Omega_s^{2}/4]+(i\delta_3-\gamma_{41})\Omega_c^{2}/4}})\}
\end{equation}

By dividing both the numerator and denominator in Eq.(\ref{R31N}) by $i(\omega+\delta_3)-\gamma_{41}$ and comparing the result with Eq.(\ref{R31}) of the three-level system, one finds that it has the same form as $R_{31}$ in the $\Lambda$-type system except that in the denominator and numerator the term $i\delta_2-\gamma_{21}$ is modified to
\begin{equation}\label{newd2g2}
i\delta_2-\gamma_{21}+\frac{\Omega_s^2}{4(i\delta_3-\gamma_{41})}
=i[\delta_2-\frac{\Omega_s^2\delta_3}{4(\delta_3^2+\gamma_{41}^2)}]-[\gamma_{21}+\frac{\Omega_s^2\gamma_{41}}{4(\delta_3^2+\gamma_{41}^2)}].
\end{equation}

In the case with $\delta_c=0$ and in the limit $\delta_s>>\delta_p$ and $\delta_s>>\gamma_{41}$, $R_{31}$ of the N-type system is similar to that of $\Lambda$-type system with the effective two-photon detuning and effective ground-state decoherence rate replaced by the relations,
\begin{equation}\label{deff}
\delta_{2,eff}\cong\delta_{2}-\frac{\Omega_s^2}{4\delta_s},
\end{equation} 
\begin{equation}\label{geff}
\gamma_{21,eff}\cong\gamma_{21}+\frac{\Omega_s^2\gamma_{41}}{4\delta_s^2}.
\end{equation} 
The physical meaning of these two relations are clear. Due to the off-resonant coupling of the control field on the transition $|2\rangle\rightarrow|4\rangle$, it introduces an ac Stark shift ($\sim-\frac{\Omega_s^2}{4\delta_3}$) on state $|2\rangle$ and an additional decoherence rate ($\frac{\Omega_s^2\gamma_{41}}{4\delta_s^2}$) on $\gamma_{21}$ due to the optical excitation and the spontaneous decay. However, one should be aware that the previous approximation of $R_{31}$ is only valid with $\delta_p<<\delta_s$. To be more precisely, we fit the EIT spectra of $D_2$ line to the complete lineshape, i.e. Eq.(\ref{Nspectrum}), to determine the parameters $\gamma_{21}$, $D$ and $\Omega_c$. Putting Eqs. (\ref{R31N}) and (\ref{Wpinifd}) into Eq.(\ref{Maxwellfd}), one can numerically calculate the output pulse after passing through a $N$-type media with arbitrary parameters.

\begin{figure}
\includegraphics[width=9cm,viewport=240 30 790 460,clip]{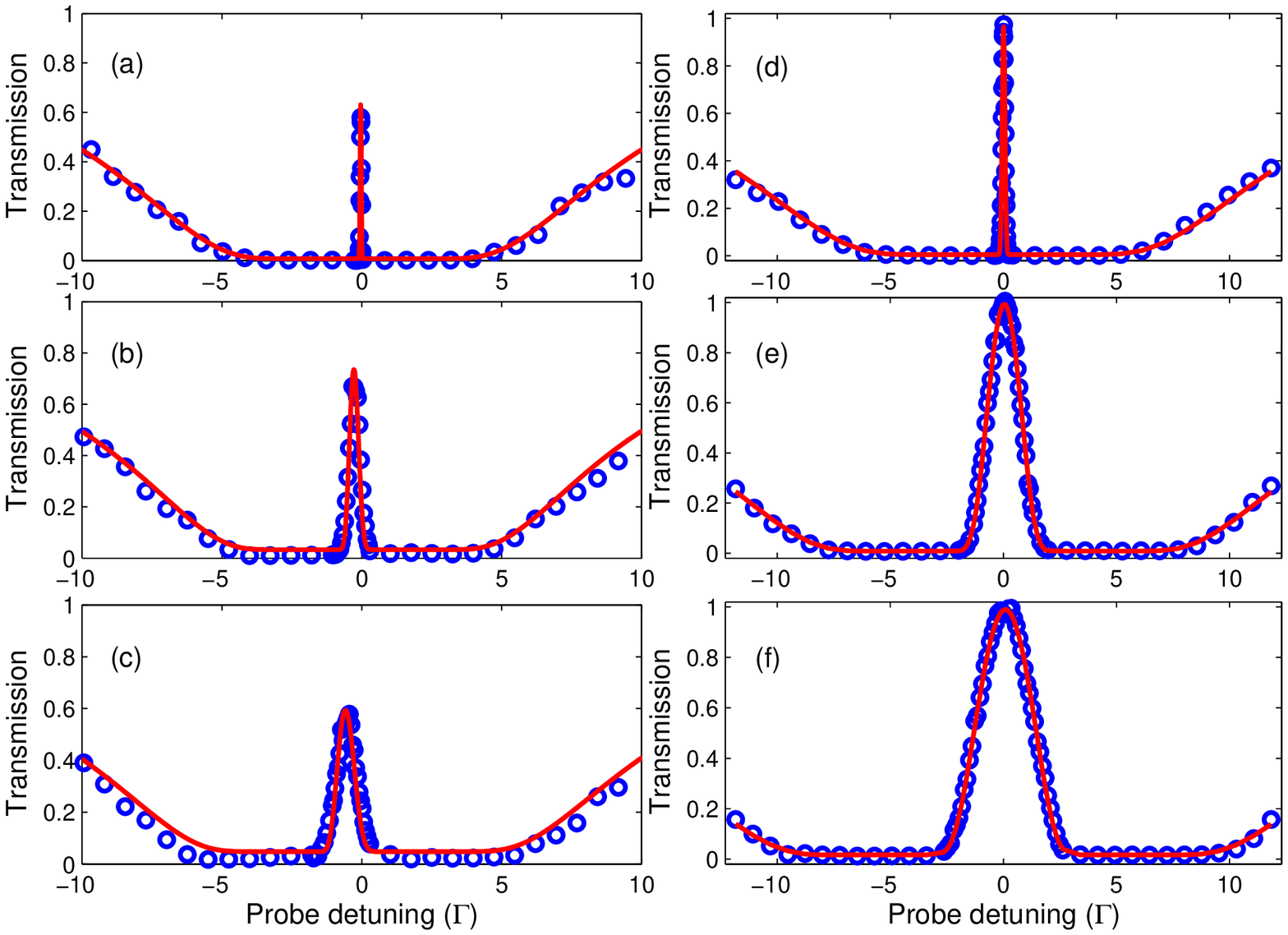}
\caption{Some representative EIT spectra for the $D_2$ ((a) to (c)) and $D_1$ ((d) to (f)) schemes. The red lines in (a)(b),and (c) are the fitting curves to Eq. (\ref{Nspectrum}) with $\gamma_{31}$=$\gamma_{41}$=0.80$\Gamma$. The fitting parameters \{$D, \gamma_{21}, \Omega_c$\} for (a) (b), and (c) are \{203(12), 0.0006(2)$\Gamma$, 1.01(2)$\Gamma$\}, \{179(16), 0.0025(16)$\Gamma$, 2.81(6)$\Gamma$\}, and \{225(25), 0.011(4)$\Gamma$, 4.10(8)$\Gamma$\}, respectively. The red lines in (d) to (f) are the fitting curves to Eq. (\ref{PowerSpec}) with $\gamma_{31}$=0.82$\Gamma$. The fitting parameters \{$D, \gamma_{21}, \delta_c, \Omega_c$\} are \{351(10), 0.00024(7)$\Gamma$, 0.0075(8)$\Gamma$, 2.05(2)$\Gamma$\}, \{399(13), 0.00039(53)$\Gamma$, 0.043(6)$\Gamma$, 7.31(6)$\Gamma$\}, and \{479(25), 0.0010(8)$\Gamma$, 0.086(11)$\Gamma$, 10.01(11)$\Gamma$\}, respectively. The quantities in the brackets are the 2$\sigma$ standard deviation of the fitting parameters.}
\end{figure}

Fig. S3 shows some representative EIT spectra for both $D_1$ and $D_2$ schemes. Two trends are clearly seen. First, the degrees of transparency in the EIT transparent peaks for the $D_1$ scheme are much higher than those in the $D_2$ scheme. Second, the EIT transparent peaks for the $D_2$ scheme have clear shifts when the control intensities are stronger. The shifts for the $D_1$ spectra are very small even for the strongest control intensity. The reason for these two trends is the same. In the $D_2$ scheme, the off-resonant excitation of the control field in the $|2\rangle\rightarrow|4\rangle$ transition is much stronger due to a relatively small detuning $\delta_s$ of 251.0916 MHz. From the fitting of the $D_1$ EIT spectra to Eq. (\ref{PowerSpec}), the parameters \{$D, \Omega_c, \gamma_{21}, \delta_c$\} can be determined. The EIT resonance shift is represented by the fitting parameter $\delta_c$. For the $D_2$ spectra, we still fit the spectra to Eq. (\ref{PowerSpec}) first just to get the parameter $\delta_c$ for resonance shifts. We then fit the same spectra to Eq. (\ref{Nspectrum}) to determine the parameters \{$D, \Omega_c, \gamma_{21}$\} by setting $\delta_c$=0. 
The systematic variations of $\gamma_{21}$ and resonance shift versus the control power are shown in Fig. 2 of the main text. The theoretical line of $\gamma_{21}$ for the $D_2$ line shown in Fig. 2 is calculated by Eq. (\ref{geff}).      

\subsection{Semi-classical calculation of the four-wave mixing} 
For the energy level illustrated in Fig. 1(b) of the main text, the off-resonant excitation of the control field on the probe transition can act as a pumping field and induces a four-wave mixing (FWM) process through absorbing the pump photons, emitting the Stoke photons, absorbing the control photons and emitting the probe photons. FWM induces a probe gain and introduces quantum noise which reduces the fidelity of a quantum memory based on EIT scheme. A better way to evaluate the effect of the FWM on EIT-based memory would be through calculation of the fidelity based on the Bloch-Langevin and Maxwell equations\cite{Fleischhauer13}. However, this calculation is very involved. Since the reduction in the fidelity is monotonically proportional to the probe gain\cite{Fleischhauer13}, we use the semi-classical Bloch-Maxwell equations to calculate the FWM probe gain to evaluate the effect of the FWM in our system. We perform the calculation both in the steady-state and pulse regime. For the pulse calculation, the complete optical Bloch equations and the Maxwell equation are numerically calculated under the assumption of perfect phase matching. The calculated FWM gain can be considered as an upper bound of the experimental results. Fig. 5(a) in the main context is based on this calculation. For the steady-state calculation, the analytic expressions are obtained under the weak probe and idler perturbation. The phase miss-matching is also considered for evaluation of the dependence of the FWM gain on the probe detuning and its intersection angle with the control beam ($\theta$)\cite{Boyer13}. The calculation is similar to that in \cite{Boyer13}, but we relieve the assumption that the Rabi frequencies for the control field on the two transitions it drives are the same. Figs. 5(b) and (c) in the main text are calculated using the steady state formula described in this section.     

Although our energy level scheme involves three levels only, a four-level model is used to facilitate our calculation. In the limit of far pump detuning (which is valid in our case with a detuning of $\sim$9.2 GHz, the basic equations for the three and four level model are the same, except for a difference in the ac-Stark shift which plays a minor role\cite{Fleischhauer13}. The energy levels and notations follow that in \cite{Boyer13}. However, it should be noted that the definition of the probe detuning $\delta_p$ in \cite{Boyer13} (and this section) differ by a negative sign to the main text and other sections in this supplementary material. Including the time dependence, the Hamiltonian of the system is
\begin{equation}\label{Hamiltonian2} 
\begin{bmatrix}
       0 & 0 & -\frac{\hbar\Omega_p^*}{2}e^{i\omega_{p}t} &   -\frac{\hbar\Omega_d^*}{2}e^{i\omega_{c}t}        \\[0.3em]
       0 & \hbar\omega_{21} &  -\frac{\hbar\Omega_c^*}{2}e^{i\omega_{c}t}           &  -\frac{\hbar\Omega_i^*}{2}e^{i\omega_{i}t}  \\[0.3em]
        -\frac{\hbar\Omega_p}{2}e^{-i\omega_{p}t}& -\frac{\hbar\Omega_c}{2}e^{-i\omega_{c}t} &          \hbar\omega_{31} & 0\\[0.3em]
 -\frac{\hbar\Omega_d}{2}e^{-i\omega_{c}t} & -\frac{\hbar\Omega_i}{2}e^{-i\omega_{i}t} & 0 & \hbar\omega_{41} 
     \end{bmatrix}.
\end{equation}
To incorporate the phase-matching issue, the position dependent phase factor should be included in the Rabi frequencies, e.g. $\Omega_p$ should be replaced by $\Omega_{p}e^{i\vec{k_p}\cdot\vec{r}}$ and so on. However, for clarity of notation, we will do that after the solution of the density matrix elements is reached.                 
Considering the unitary transformation 
\begin{equation}
T=\begin{bmatrix}
1 & 0 &0 &0\\
0 & e^{i(\omega_p-\omega_c)t} & 0 &0\\
0 & 0 & e^{i\omega_{p}t} & 0\\
0 & 0 & 0 & e^{i\omega_{c}t}
\end{bmatrix},
\end{equation}
the Hamiltonian in the new basis is $H'=THT^{\dagger}+i\hbar\frac{dT}{dt}T^{\dagger}$. To completely eliminate the explicit time independence in the Hamiltonian, the frequencies of the control, probe and idler need to satisfy 
\begin{equation}  
2\omega_c=\omega_p+\omega_i,       
\end{equation}
which is the energy conversation condition of the FWM process. The Hamiltonian in the new basis then becomes
\begin{equation}
H=\hbar\begin{bmatrix}
0 & 0 & -\frac{\Omega_p^*}{2}&-\frac{\Omega_d^*}{2}\\
0 & -\delta_2 & -\frac{\Omega_c^*}{2} &-\frac{\Omega_i^*}{2}\\
-\frac{\Omega_p}{2} & -\frac{\Omega_c}{2} & -\delta_p & 0\\
-\frac{\Omega_d}{2} & -\frac{\Omega_i}{2} & 0 & -\delta_d
\end{bmatrix}.
\end{equation}
The evolution equations for the density matrix elements are,
\begin{eqnarray}
\dot{\sigma_{11}}=\frac{i}{2}(\Omega_p^*\sigma_{31}+\Omega_d^*\sigma_{41}-\Omega_p\sigma_{13}-\Omega_d\sigma_{14})+\Gamma_{31}\sigma_{33}+\Gamma_{41}\sigma_{44},\\
\dot{\sigma_{22}}=\frac{i}{2}(\Omega_c^*\sigma_{32}+\Omega_i^*\sigma_{42}-\Omega_c\sigma_{23}-\Omega_i\sigma_{24})+\Gamma_{32}\sigma_{33}+\Gamma_{42}\sigma_{44},\\
\dot{\sigma_{33}}=\frac{i}{2}(\Omega_p\sigma_{13}+\Omega_c\sigma_{23}-\Omega_p^*\sigma_{31}-\Omega_c^*\sigma_{32})-\Gamma_{3}\sigma_{33},\\
\dot{\sigma_{44}}=\frac{i}{2}(\Omega_d\sigma_{14}+\Omega_i\sigma_{24}-\Omega_d^*\sigma_{41}-\Omega_i^*\sigma_{42})-\Gamma_{4}\sigma_{44},\\
\dot{\sigma_{21}}=\xi_{21}\sigma_{21}+\frac{i}{2}(\Omega_c^*\sigma_{31}+\Omega_i^*\sigma_{41}-\Omega_p\sigma_{23}-\Omega_d\sigma_{24}),\\
\dot{\sigma_{31}}=\xi_{31}\sigma_{31}+\frac{i}{2}[\Omega_p(\sigma_{11}-\sigma_{33})+\Omega_c\sigma_{21}-\Omega_d\sigma_{34}],\\
\dot{\sigma_{32}}=\xi_{32}\sigma_{32}+\frac{i}{2}[\Omega_p\sigma_{12}+\Omega_c(\sigma_{22}-\sigma_{33})-\Omega_i\sigma_{34}],\\
\dot{\sigma_{41}}=\xi_{41}\sigma_{41}+\frac{i}{2}[\Omega_d(\sigma_{11}-\sigma_{44})+\Omega_i\sigma_{21}-\Omega_p\sigma_{43}],\\
\dot{\sigma_{42}}=\xi_{42}\sigma_{42}+\frac{i}{2}[\Omega_i(\sigma_{22}-\sigma_{44})+\Omega_d\sigma_{12}-\Omega_c\sigma_{43}],\\
\dot{\sigma_{43}}=\xi_{43}\sigma_{43}+\frac{i}{2}(\Omega_d\sigma_{13}+\Omega_i\sigma_{23}-\Omega_p^*\sigma_{41}-\Omega_c^*\sigma_{42}),\\
\end{eqnarray}
where $\xi_{21}=i(\delta_p-\delta_c)-\gamma_{21}=i\delta_2-\gamma_{21}, \xi_{31}=i\delta_p-\gamma_{31}, \xi_{32}=i\delta_c-\gamma_{32}, \xi_{41}=i\delta_d-\gamma_{41}, \xi_{42}=i(\delta_d-\delta_2)-\gamma_{42}, \xi_{43}=i(\delta_d-\delta_p)-\gamma_{43}, \Gamma_3=\Gamma_{31}+\Gamma_{32}, \Gamma_4=\Gamma_{41}+\Gamma_{42}$.
We consider the weak probe and idler perturbation and calculate the steady-state solution. To the zero order (i.e. $\Omega_p=0=\Omega_i$), the solutions of the density matrix elements $\sigma_{12}, \sigma_{13}, \sigma_{24}, \sigma_{43}$ and their transpose are zero, because there are no laser fields to create their coherences. By solving the two coherence evolution equations for $\sigma_{32}$ and $\sigma_{41}$ and the four population equations, in addition to the population conservation law, on obtains the steady-state solution for the zero-order  populations, which are   
\begin{eqnarray}
\sigma_{11}^{(0)}=\frac{\frac{\Gamma_{42}}{\Gamma_{31}}(1+\frac{2\Gamma_3|\xi_{32}|^2}{\gamma_{32}|\Omega_c|^2})}{2(1+\frac{\Gamma_{42}}{\Gamma_{31}}+\frac{\Gamma_{42}\Gamma_3|\xi_{32}|^2}{\Gamma_{31}\gamma_{32}|\Omega_c|^2}+\frac{\Gamma_4|\xi_{41}|^2}{\gamma_{41}|\Omega_d|^2})},\\
\sigma_{22}^{(0)}=\frac{1+\frac{2\Gamma_4|\xi_{41}|^2}{\gamma_{41}|\Omega_d|^2}}{2(1+\frac{\Gamma_{42}}{\Gamma_{31}}+\frac{\Gamma_{42}\Gamma_3|\xi_{32}|^2}{\Gamma_{31}\gamma_{32}|\Omega_c|^2}+\frac{\Gamma_4|\xi_{41}|^2}{\gamma_{41}|\Omega_d|^2})},\\
\sigma_{33}^{(0)}=\frac{\frac{\Gamma_{42}}{\Gamma_{31}}}{2(1+\frac{\Gamma_{42}}{\Gamma_{31}}+\frac{\Gamma_{42}\Gamma_3|\xi_{32}|^2}{\Gamma_{31}\gamma_{32}|\Omega_c|^2}+\frac{\Gamma_4|\xi_{41}|^2}{\gamma_{41}|\Omega_d|^2})},\\
\sigma_{44}^{(0)}=\frac{1}{2(1+\frac{\Gamma_{42}}{\Gamma_{31}}+\frac{\Gamma_{42}\Gamma_3|\xi_{32}|^2}{\Gamma_{31}\gamma_{32}|\Omega_c|^2}+\frac{\Gamma_4|\xi_{41}|^2}{\gamma_{41}|\Omega_d|^2})}.\\
\end{eqnarray}
Putting the zero-order solutions into the four first-order equations for the coherences, $\sigma_{21}$, $\sigma_{31}$, $\sigma_{42}$, and $\sigma_{43}$, one obtains the steady-state, and first-order solutions of the two coherence terms, $\sigma_{31}$ and $\sigma_{42}$. Now, we restore the position-dependent phase factors into the Rabi frequency. The relation between the atomic coherence and the slowly-varying macroscopic polarization is
\begin{equation}
P_{p(i)}=2n_{a}d_{31(42)}\sigma_{31(42)}e^{-i\vec{k_{p(i)}}\cdot\vec{r}},
\end{equation}
where $n_a$ is the atomic density. The slowly-varying polarizations can be written as
\begin{eqnarray}
P_p=\varepsilon_0\chi_{pp}E_{p}+\varepsilon_0\chi_{pi}E_{i}^*e^{i(2\vec{k_c}-\vec{k_i}-\vec{k_p})\cdot\vec{r}},\\
P_i=\varepsilon_0\chi_{ii}E_{i}+\varepsilon_0\chi_{ip}E_{p}^*e^{i(2\vec{k_c}-\vec{k_p}-\vec{k_i})\cdot\vec{r}},
\end{eqnarray} 
with the susceptibilites 
\begin{equation}
\chi_{pp}=\frac{in_a|d_{31}|^2}{D\varepsilon_0\hbar}\{\frac{\xi_{43}^*\xi_{42}^*+\frac{1}{4}|\Omega_c|^2(1-|\epsilon|^2)}{\xi_{32}^*}\sigma_{22,33}^{(0)}-[\frac{\xi_{42}^*\xi_{21}\xi_{43}^*}{|\Omega_c^2|/4}+(|\epsilon|^2\xi_{43}^*+\xi_{21})]\sigma_{11,33}^{(0)}
\\+|\epsilon|^2\frac{\xi_{21}\xi_{42}^*+\frac{1}{4}|\Omega_c|^2(|\epsilon|^2-1)}{\xi_{41}^*}\sigma_{11,44}^{(0)}\},
\end{equation}
\begin{equation}
\chi_{pi}=\frac{in_{a}d_{31}d_{42}\Omega_c\Omega_d}{\varepsilon_0\hbar|\Omega_c|^2D}\{\frac{\xi_{21}\xi_{42}^*+\frac{1}{4}|\Omega_c|^2(|\epsilon|^2-1)}{\xi_{32}}\sigma_{22,33}^{(0)}+(\xi_{43}^*+\xi_{21})\sigma_{22,44}^{(0)}+\frac{\xi_{43}^*\xi_{42}^*}{\xi_{41}}\sigma_{11,44}^{(0)}\},
\end{equation}
\begin{equation}
\chi_{ii}=\frac{in_a|d_{42}|^2\xi_{31}^*}{\varepsilon_{0}\hbar D}\{|\epsilon|^2\frac{4\xi_{43}\xi_{31}^*+|\Omega_c|^2(|\epsilon|^2-1)}{\xi_{41}^*\xi_{31}^*}\sigma_{11,44}^{(0)}+\frac{4\xi_{21}^*\xi_{31}^*+|\Omega_c^2|^2(1-|\epsilon|^2)}{4\xi_{32}^*\xi_{31}^*}\sigma_{22,33}^{(0)}-[\frac{\xi_{21}^*\xi_{43}}{|\Omega_c|^2/4}+\frac{\xi_{43}+\xi_{21}^*|\epsilon|^2}{\xi_{31}^*}]\sigma_{22,44}^{(0)}\},
\end{equation}
\begin{equation}
\chi_{ip}=\frac{in_{a}d_{31}d_{42}\xi_{31}^*\Omega_c\Omega_d}{\varepsilon_0\hbar D|\Omega_c|^2}\{[\frac{\xi_{43}}{\xi_{32}}-\frac{|\Omega_c|^2(1-|\epsilon|^2)}{4\xi_{32}\xi_{31}^*}]\sigma_{22,33}^{(0)}+\frac{\xi_{43}+\xi_{21}^*}{\xi_{31}^*}\sigma_{11,33}^{(0)}+[\frac{\xi_{21}^*}{\xi_{41}}-\frac{|\Omega_c|^2(|\epsilon|^2-1)}{4\xi_{41}\xi_{31}^*}]\sigma_{11,44}^{(0)}\}, ,
\end{equation}
\begin{equation}
D=\frac{\xi_{31}\xi_{42}^*\xi_{21}\xi_{43}^*}{|\Omega_c|^2/4}+\xi_{43}^*(\xi_{42}^*+|\epsilon|^2\xi_{31})+\xi_{21}(\xi_{42}^*|\epsilon|^2+\xi_{31})+\frac{1}{4}(|\epsilon|^2-1)^2,
\end{equation}
where $\epsilon=\Omega_d/\Omega_c$ and in the scheme of Fig. 1(b) of the main text, $\epsilon=-\sqrt{7}$ and $\sigma_{ii,jj}^{(0)}=\sigma_{jj}^{(0)}-\sigma_{ii}^{(0)}$ ($i=\{1,2\}, j=\{3,4\}$) is the population difference. For the beams nearly copropagating on the z-axis, the steady-state Maxwell equations for the probe and idler fields are
\begin{eqnarray}
\frac{\partial E_p}{\partial z}=\frac{ik_p}{2}\chi_{pp}E_p+\frac{ik_p}{2}\chi_{pi}e^{i\Delta k_{z}z}E_i^*,\\
\frac{\partial E_i}{\partial z}=\frac{ik_i}{2}\chi_{ii}E_i+\frac{ik_i}{2}\chi_{ip}e^{i\Delta k_{z}z}E_p^*,
\end{eqnarray}
where $\Delta k_{z}=2n_c\vec{k_c}-k_{p}cos\theta-k_{i}sin\theta$. The index of refraction for the control field $n_c$ is explicitly added since the control field off-resonantly drives the probe transition with the major population located in state $|1\rangle$. We do not consider the Maxwell equation for the control field. The index of refraction $n_c$ is calculated from the two-level response\cite{Boyer13}, i.e. $n_c=\sqrt{\chi_c}\cong 1+\chi_c/2$ with
\begin{equation}
\chi_c=-\frac{n_ad_{41}^2}{\varepsilon_0\hbar}\frac{\delta_d}{\delta_d^2+\Gamma_4^2/4}.
\end{equation}
The coupled Maxwell equations for the probe and idler fields can be solved to be
\begin{equation}\label{Epg}
E_p=E_{p0}exp(\delta aL)[cosh(\xi L)+\frac{a}{\xi}sinh(\xi L)],
\end{equation}
\begin{equation}
E_i^*=E_{p0}\frac{a_{ip}}{\xi}exp(\delta aL)sinh(\xi L),
\end{equation}
where $a_{pj}=\frac{ik_p}{2}\chi_{pj}, a_{ij}=\frac{ik_i}{2}\chi_{ij}, j=\{i,p\}$ and $a=(a_{ii}+a_{pp}-i\Delta k_z)/2, \delta a=(a_{pp}-a_{ii}+i\Delta k_z)/2, \xi=\sqrt{a^2-a_{pi}a_{ip}}$. The FWM probe gain can be calculated by Eq. (\ref{Epg}).

\subsection{Experimental test of the four-wave mixing gain} 
\begin{figure}
\includegraphics[width=12cm,viewport=200 0 750 400,clip]{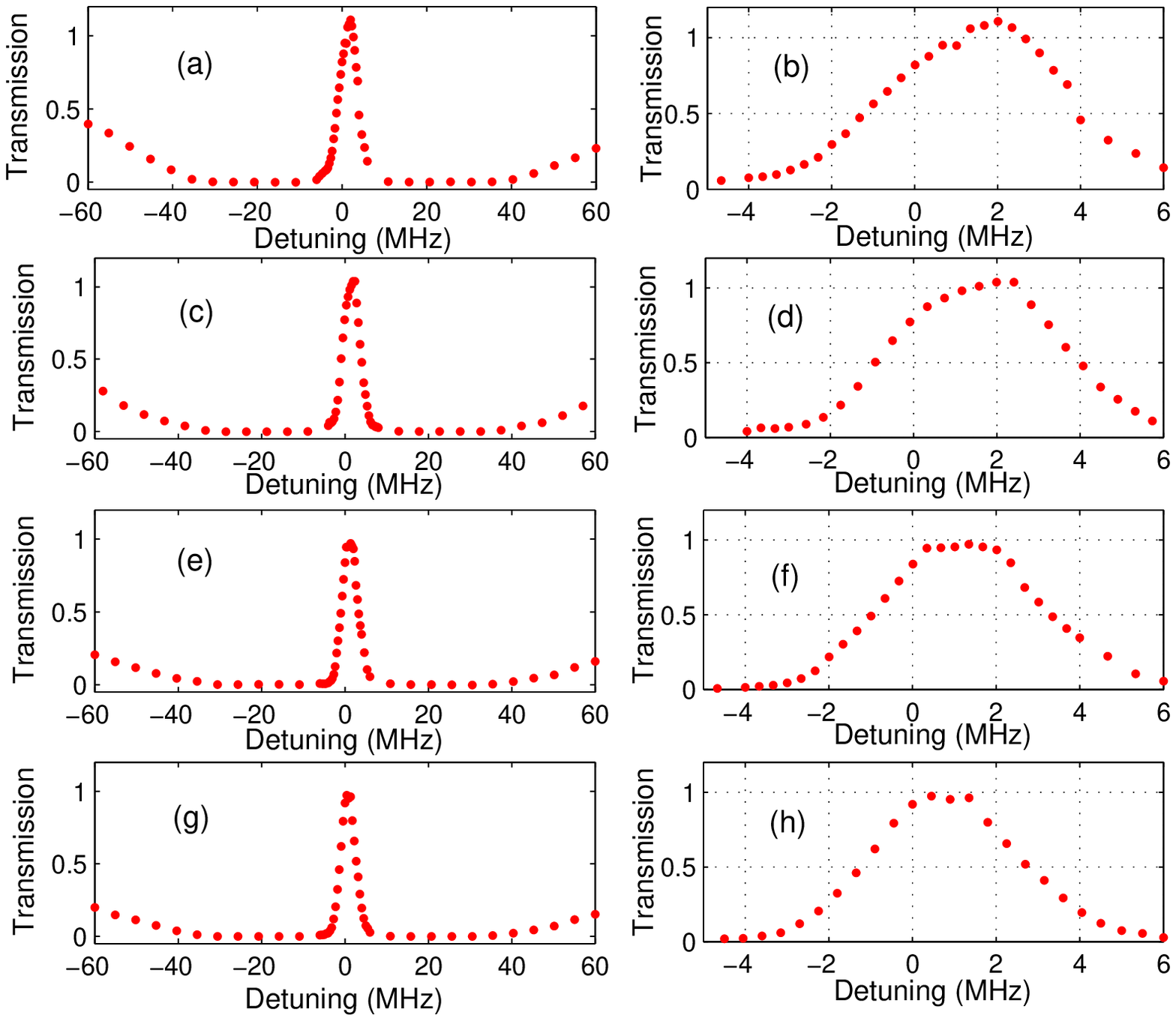}
\caption{Some representative spectra for various pump detuning. The detunings are 1.5, 1.77, 4, and 9.192 GHz for (a),(c),(e), and (f), respectively. The figures in the right column are the zoom-in for the central transparent peak of those in the left column.}  
\end{figure}

Based on the theoretical calculation, the four-wave mixing gain of the probe due to the off-resonant excitation of the control field on the probe transition is less than 2\% even with an OD of up to 1000, as described in the main text. Such a small gain is within the experimental uncertainty of the transmission determination of the slow light pulse or the transmission of the steady-state EIT spectrum. However, such an estimation of FWM gain is solely based on the theoretical calculation. To provide an experimental support of these calculations, we have performed an additional experiment on the four-wave mixing gain of the probe. We add one pump beam which is coupled into the same fiber for the control beam. Therefore, this pump beam completely overlaps with the control beam. For this pump beam, its detuning relative to the probe transition is tunable. We could adjust its detuning to a small value such that the FWM gain introduced by this pump beam is large and measurable. We then study the asymptotic behavior of the FWM gain versus its detuning. At the detuning of 9.192 GHz, the frequency of the pump field is the same as that of the control field and its role is the same as the control field. This allows us to quantitatively estimate the FWM gain due to the off-resonant excitation of the control field on the probe transition based on the experimental basis.

At an OD of 603 and an $\Omega_c$ of 6.2 $\Gamma$ corresponding to $\zeta$=2.7 for the slow light, we measure the steady-state EIT spectra for various pump detuning. The angle between the probe and control beam is $0.5^0$. Under the presence of the pump field, we found that the population can be optically pumped out of the system (e.g. to $|F=4, m=4\rangle$) such that the OD may be smaller for a smaller pump detuning. To minimize such a population loss, we turn on the pump beam only 3 $\mu$s before the probe beam is on. With such an arrangement, the population loss due to the pump beam is kept less than 10\% for a pump detuning of lager than 2 GHz.    
 
Limited by the available power, the maximum power of the pump beam is 7.7 mW. This power is about half of the control power when $\zeta$ is kept at 2.7 for an OD of 603, as in the case for Fig. 5(d) of the main text. Figure S4 shows some representative spectra for various pump detuning. The left column shows the full range spectra while the right column shows the zoom-in of the central transparent peaks. With the smaller pump detuning, the four-wave mixing gains are clearly seen. The FWM gain is determined by the peak probe transmission to that without the presence of the pump beam. With a pump detuning of 9.192 GHz, the FWM gain is $\lesssim$ 0.6\%, as shown in Fig. 5(d) of the main text. Based on this result and the scaling law of FWM gain with OD and pump power, we estimate that the FWM gain introduced by the control field at an OD of 816, as the maximum OD in Fig. 4(d), is within $\sim 3\%$. This provides an experimental check of the effect of FWM in our experiment.
